\documentclass{ws-ijmpc}

\begin{document}

\markboth{V. Dexheimer et al}
{Instructions for Typing Manuscripts (Paper's Title)}

\catchline{}{}{}{}{}

\title{PROTO-NEUTRON AND NEUTRON STARS}

\author{V. Dexheimer}

\address{FIAS,
Johann Wolfgang Goethe University,
Ruth-Moufang-Str. 1,
60438 Frankfurt am Main,
Germany
\\dexheimer@th.physik.uni-frankfurt.de}

\author{S. Schramm}

\address{CSC, FIAS, Johann Wolfgang Goethe-University, Max-von-Laue-Straße 1,
60438 Frankfurt am Main, Germany}

\author{H. Stoecker}

\address{FIAS, ITP, Johann Wolfgang Goethe University,
Ruth-Moufang-Str. 1,
60438 Frankfurt am Main,
Germany}

\maketitle

\begin{history}
\received{Day Month Year}
\revised{Day Month Year}
\end{history}

\begin{abstract}
 The parity doublet model, containing the SU(2) multiplets including the baryons identified as the
 chiral partners of the nucleons is applied to neutron stars. The maximum mass for the star is calculated for different stages of
 the cooling taking into account finite temperature/entropy effect, trapped neutrinos and fixed baryon number. Rotation effects are also included.

\end{abstract}

\ccode{PACS Nos.: 11.25.Hf, 123.1K}

\section{Introduction}

It has been shown that effective chiral hadronic models can
satisfactorily describe nuclear matter, properties of finite
nuclei as well as the structure of neutron stars\cite{model}.
While most of the approaches are based on variations of the linear
sigma model or its non-linear realization, in the approach
discussed here the scalar sigma meson serves to split the nucleon
and its chiral partner (particle with opposite parity), while in
the chirally restored phase both baryonic states become degenerate
but not massless. The presence of the chiral partner allows to
have a bare mass term $m_0$ in the Lagrangian density in such a
way that it does not break chirality because the physical fields
in this case are a mixture of the fields of the particles and
their chiral partners\cite{kunihiro,jido,ZParity}.

\section{The parity Doublet Model}

It is assumed that the star is in chemical equilibrium and the baryons interact through the mesons $\sigma$, $\omega$ and $\rho$ (included in order to reproduce the high asymmetry between neutrons and protons). Electrons are included to insure charge neutrality. The Lagrangian density in mean field approximation becomes
\begin{eqnarray}
L_{MFT}=L_{kin}+L_{Bscal}+L_{Bvec}+L_{vec}+L_{scal}+L_{SB},
\end{eqnarray}
 \begin{eqnarray}
L_{Bscal}+L_{Bvec}=-\sum_i \bar{\psi_i}[g_{i\omega}\gamma_0\omega^0+g_{i\rho}\gamma_0\tau_3\rho^0+M_i^*]\psi_i,
\end{eqnarray}
\begin{eqnarray}
L_{vec}=-\frac{1}{2}(m_\omega^2\omega^2+m_\rho^2\rho^2)-g_4[\omega^4+6\rho^2\omega^2+\rho^4],
\end{eqnarray}
\begin{eqnarray}
L_{scal}=\frac{1}{2}\mu^2\sigma^2-\frac{\lambda}{4}\sigma^4,
\end{eqnarray}
\begin{eqnarray}
L_{SB}=m_\pi^ 2 f_\pi\sigma\nonumber,
\end{eqnarray}
where besides the kinetic term for the fermions there are terms of interaction between the baryons and the scalar and vector mesons, self-interaction terms
for the vector and scalar mesons and an explicit symmetry breaking term included in order to reproduce the masses of the pseudo-scalar mesons.

The effective masses of the nucleons and their chiral partners reproduce their measured values at low densities, when the scalar condensate has its vacuum value, and they go to a specific value $m_0=790$ MeV (chosen in order to have a physical compressibility at saturation\cite{artigonovo}) at high densities, when the scalar condensate go to zero
\begin{eqnarray}
M^*_\pm=\sqrt{\left[\frac{(M_{N_+}+M_{N_-})^2}{4}-m_0^2\right]\frac{\sigma^2}{\sigma_0^2}+m_0^2}\pm\frac{M_{N_+}-M_{N_-}}{2}\frac{\sigma}{\sigma_0}.
\end{eqnarray}

A possible, but not definite, candidate for the nucleon chiral
partner is the $N_-$(1535), but besides that, the case with
$N_-$(1200) is included to study the effect of the mass of the
$N_-$. This variation has drastic consequences on the results.
For this model, four different cases first studied in
\cite{ZParity} are applied to proto-neutron and neutron stars:
\begin{itemize}
\item P1: $M_{N_-}=1200$MeV and $g_4=0$
\item P2: $M_{N_-}=1200$MeV and $g_4=200$
\item P3: $M_{N_-}=1500$MeV and $g_4=0$
\item P4: $M_{N_-}=1500$MeV and $g_4=200$
\end{itemize}

As will be shown, the results are also very sensitive to the value
of the coupling constant of the self-interaction of the vector
mesons $g_4$.

\section{Neutron Stars}

\begin{figure}
\begin{center}
\includegraphics[width=0.5\textwidth]{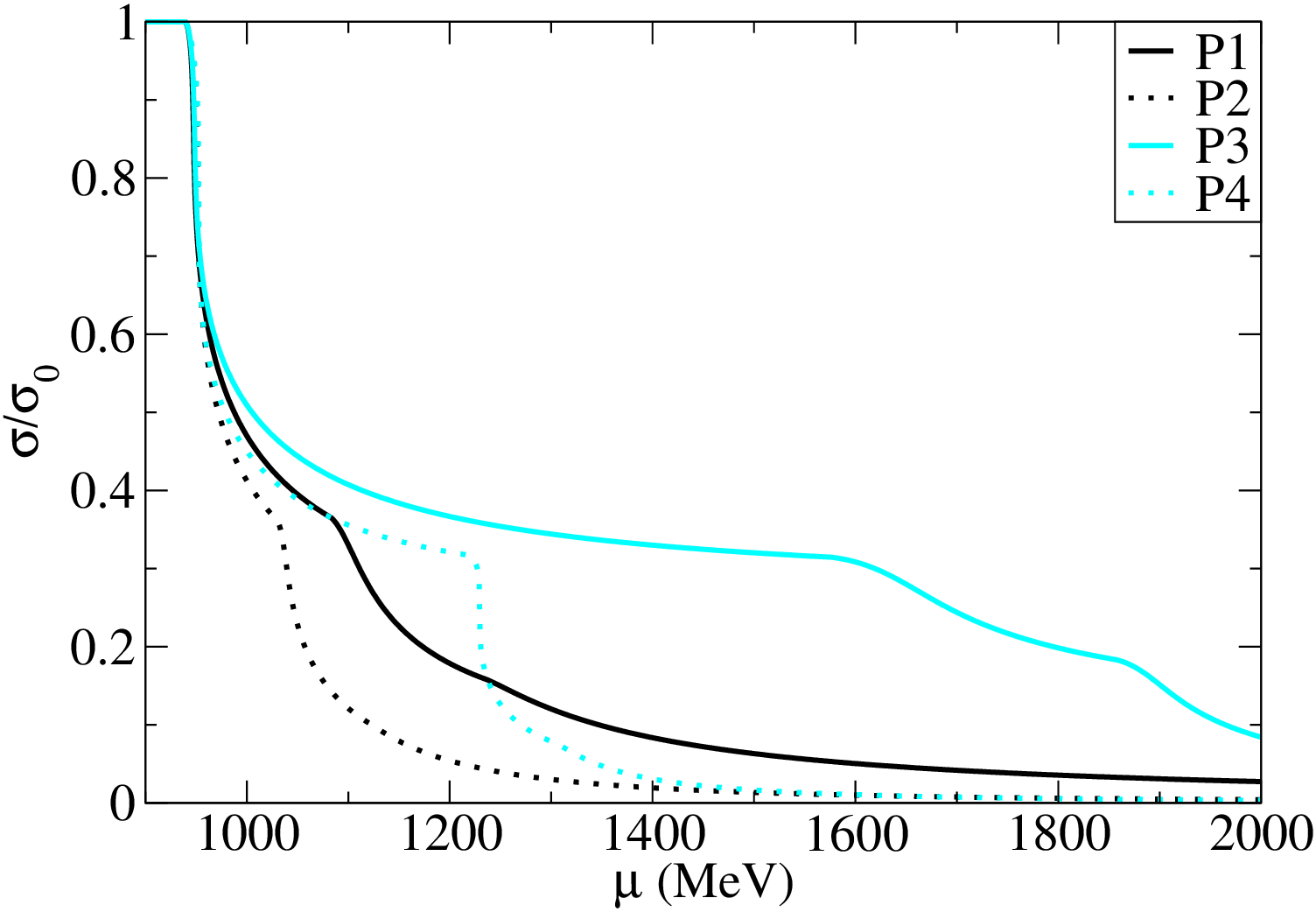}
\end{center}
\caption{Scalar condensate as a function of chemical potential for different configurations}
\label{trans}
\end{figure}

For the parity model, when the density increases, the protons, the
neutron and the proton chiral partners appear, respectively. Since
the matter is not symmetric, in general the two isospin states of
the chiral partner appear at different densities, which induces
the chiral restoration to happen earlier compared with the
symmetric nuclear matter case. As can be seen from Fig.\ref{trans}
the scalar condensate used in this case as the order parameter for
the chiral restoration shows that this transition happens very
smoothly, and can be identified as a crossover for all cases. This
effect is caused by the introduction of beta equilibrium and
charge neutrality. It can also be seen in this plot that for $P3$
the chiral restoration caused by the appearance of the chiral
partners happens at very high densities, beyond the ones possible
inside a neutron star.

\begin{figure}
\begin{center}
\includegraphics[width=0.5\textwidth]{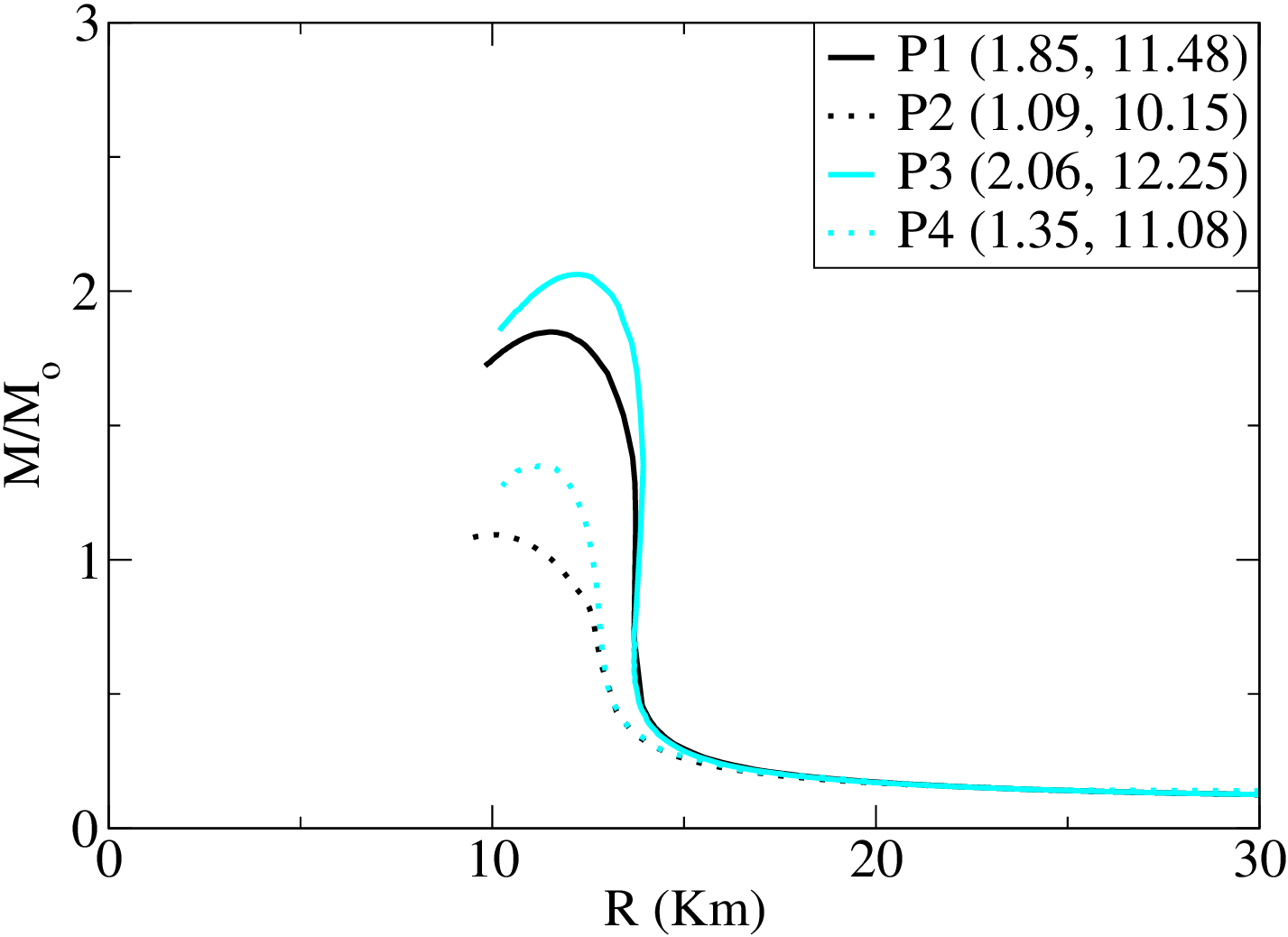}
\end{center}
\caption{Star mass as a function of radius for different configurations}
\label{massp}
\end{figure}

The maximum mass of the star is calculated solving the TOV equations and it shows that higher maximum values are reached for $g_4$ equal to zero (Fig.\ref{massp}). That means that when $g_4$ increases, the value of the vector meson $\omega$ related to it decreases its value and consequently its repulsive effect, in such a way that the star can hold smaller quantity of mass against collapse. Besides the EOS for the dense part of the star, the EOS for an outer crust, an inner crust and an atmosphere have been added\cite{crust}.

\section{Cooling}

To study the cooling of such complex system two different features
are taken into account separately: finite entropy per baryon and
lepton number. The reason for considering finite entropy instead
of temperature comes from the fact that in this case the
temperature is higher in the center of the star and colder at the
edge, which is more realistic than assuming the whole star at
equal temperature. The maximum mass of the star is higher for
higher entropies because the thermal effects make the EOS stiffer.

Trapped neutrinos with a chemical potential $\mu_\nu$ are included by fixing the
lepton number defined as $Y_l=(\rho_e+\rho_{\nu_e})/\rho_B$. In consequence
there will be a large number of neutrinos in the star but also an increased
electron density. Therefore, demanding charge neutrality, the proton density increases. For the configurations with $g_4\neq0$, the high proton fraction delays the appearance of the proton chiral partner and for having less degrees of freedom the EOS becomes stiffer. For the configurations with $g_4=0$, the chiral partners appears smoothly not having much effect, so the only effect in this case is that the high proton fraction makes the star more isospin symmetric and thus the Fermi energy smaller and the EOS softer.

\begin{figure}
\begin{center}
\includegraphics[width=0.49\textwidth]{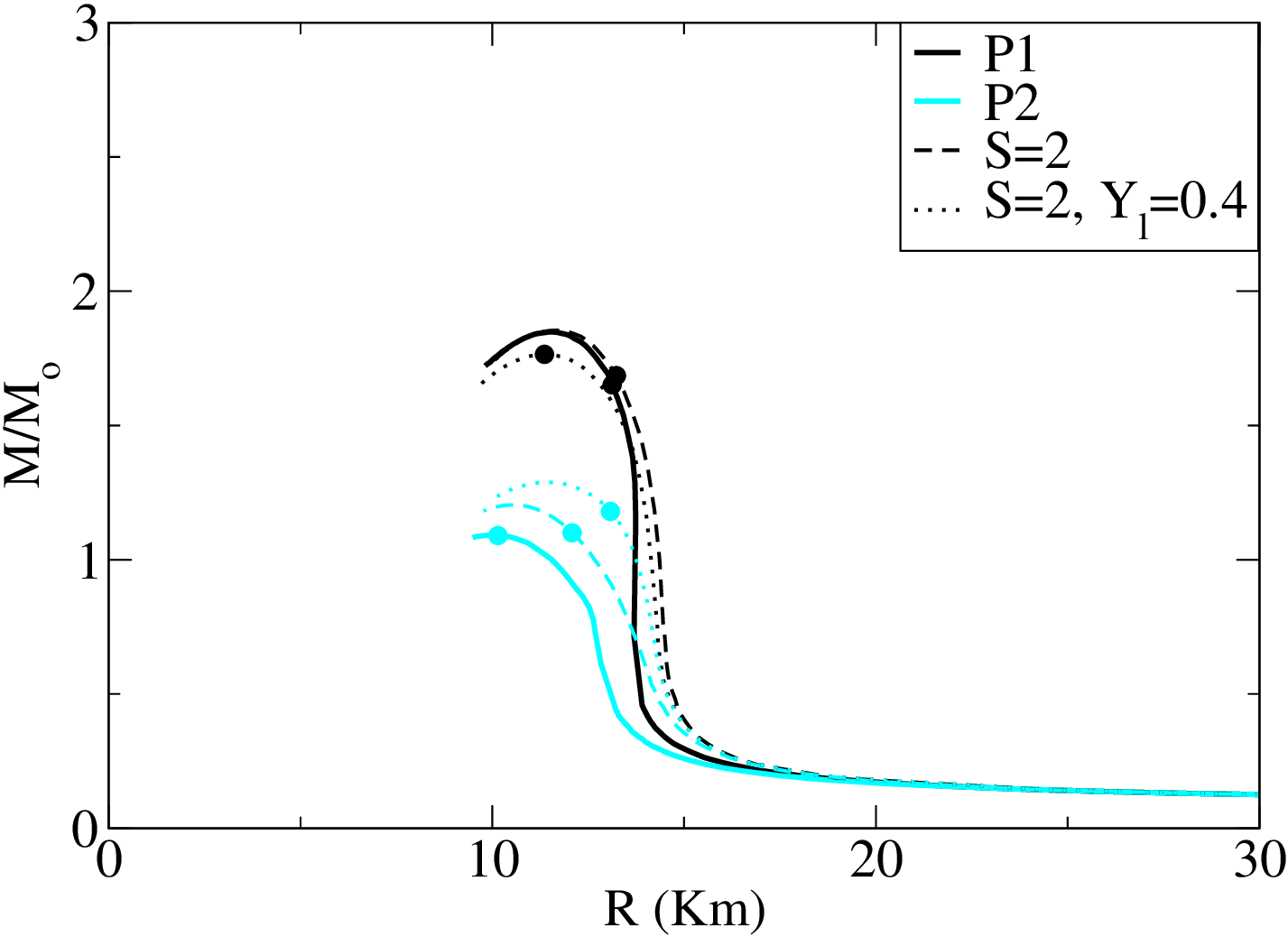}
\end{center}
\caption{Star mass as a function of radius for different stages of the cooling for the configurations with $M_{N_-}=1200$ MeV}
\label{cooling}
\end{figure}

\begin{figure}
\begin{center}
\includegraphics[width=0.49\textwidth]{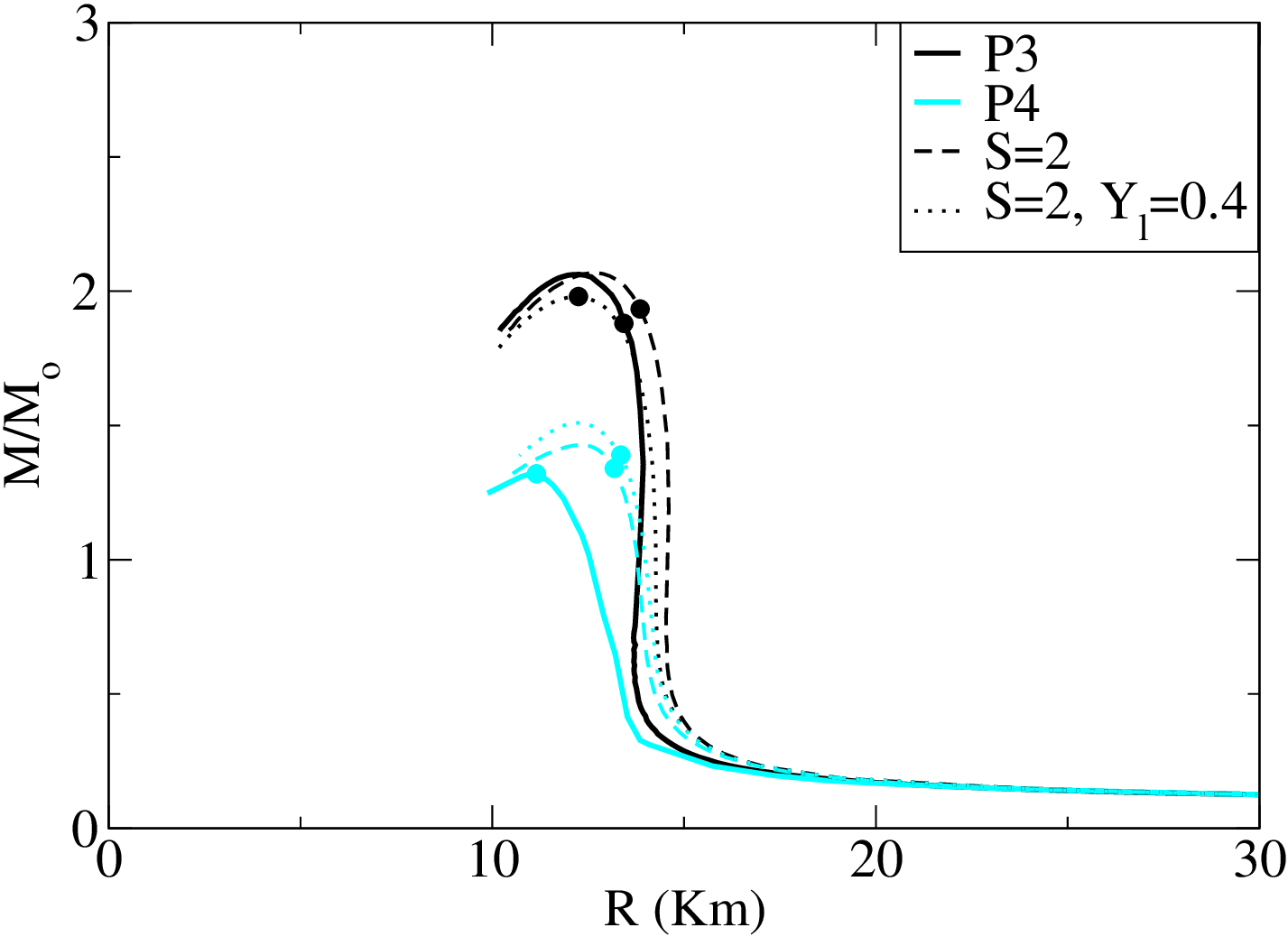}
\end{center}
\caption{Star mass as a function of radius for different stages of the cooling for the configurations with $M_{N_-}=1500$ MeV}
\label{cooling2}
\end{figure}

These two features can be put together to describe the evolution of the star. After the supernova explosion,
the star is still warm so the entropy per baryon is fixed to $S=2$ (the temperature increases from $0$ at the edge up to $45$ MeV in the center).
The star still contains a high abundance of neutrinos that were trapped during the explosion so the lepton
number is fixed to $Y_l=0.4$. After 10 to 20 seconds the neutrinos can escape and beta equilibrium is established. After about one minute the temperature of the star has dropped below 1 MeV and the entropy per baryon can be considered zero.
As it can be seen in Fig.\ref{cooling} and Fig.\ref{cooling2}, the effect of entropy and lepton number together is complicated. In the cases with finite $g_4$ (P2 and P4) the  star's maximum mass decreases with time while in the case with $g_4$ equal to zero (P1 and P3), the star's maximum mass increases with time.

The problem is that this calculation does not take into account that the baryon number cannot change with time, otherwise the star would collapse into a black hole\cite{baryonmass}. If the baryon number is fixed the maximum masses for every configuration decrease with time. This maximum masses with fixed baryon numbers are represented with dots in Fig.\ref{cooling} and Fig.\ref{cooling2}.

\section{Rotation}

\begin{figure}
\begin{center}
\includegraphics[width=0.5\textwidth, clip,trim= 0 0 0 0]{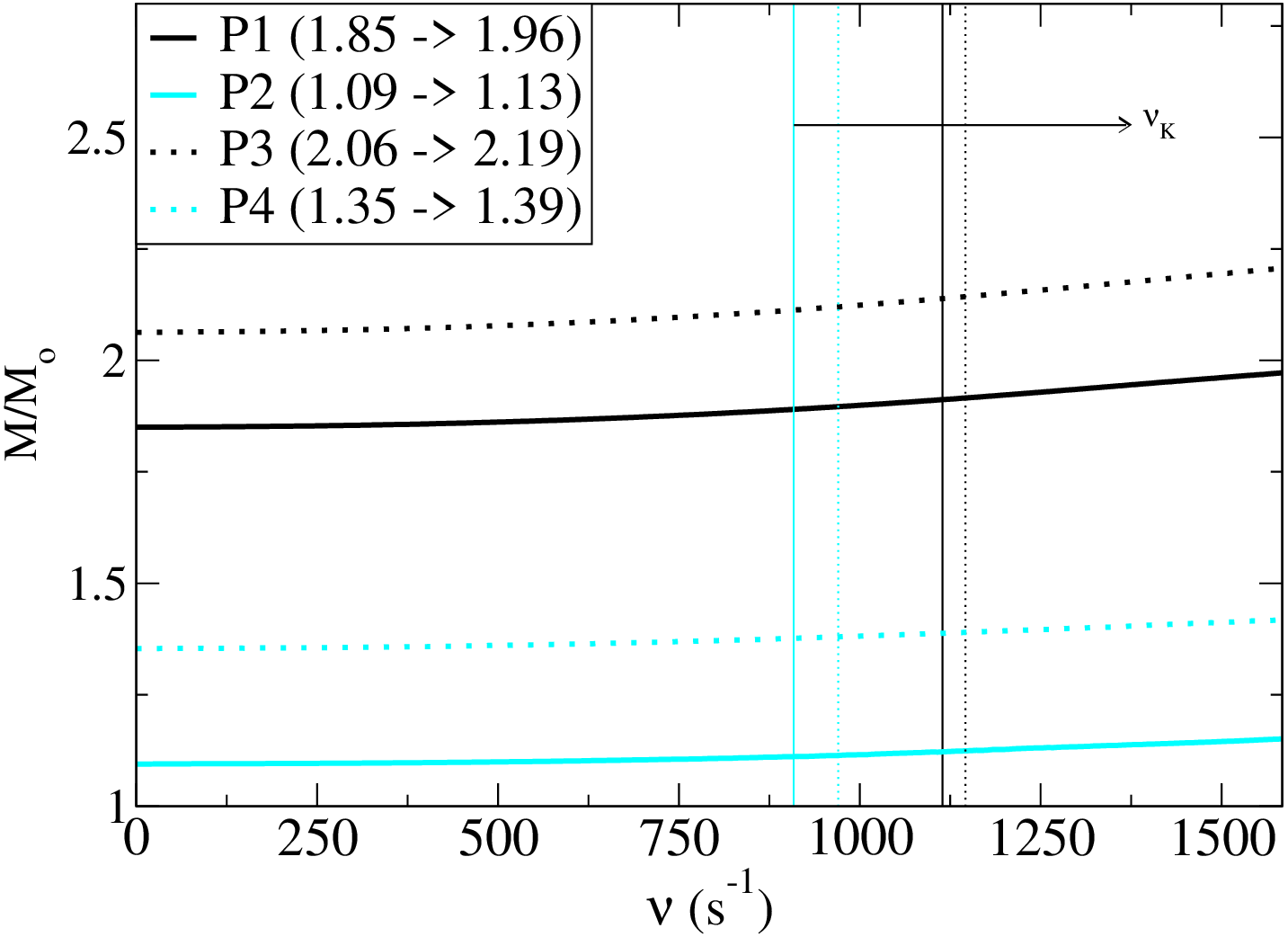}
\end{center}
\caption{Star mass as a function of rotational frequency for different configurations}
\label{rot}
\end{figure}

The maximum frequency with which a star can rotate without
starting to expel matter on the equator, named Kepler frequency,
has been determined by including monopole and quadrupole
corrections to the metric due to the rotation and solving the
self-consistency equation for $\Omega_K$ as it was derived in
\cite{glendenning}. The rotation of the star generates a
modification of the metric and the higher the rotational
frequency, the higher is the mass and radius of the
star\cite{neutronSchramm2}. For the different sets of parameters
the increase in the maximum masses of the stars from $\nu=0$ to
$\nu=\nu_K$ fixing the baryon number to the value for zero
frequency is of smaller than $5\%$ (Fig. \ref{rot}), different
from the $15\%$ mass increase for the case that the baryon number
is not fixed. But this situation can be identified as the spin
down of a cold star with a certain baryon number that continues
until it emits all its energy and stops rotating. At this point
the star is considered ``dead''.

\section{Conclusion}

With increasing density .i.e. toward the center of the star,
chiral partners appear, reaching a point where they exist with
about the same densities as the corresponding nucleons. The
decrease in the scalar condensate signals the restoration of the
chiral phase. This transition is a cross over for any of the
studied configurations due to the requirements of beta equilibrium
and charge neutrality.

The maximum mass of the star is higher when the coupling constant
$g_4$ is set to zero (P1 and P3) although these values have to be
constrained to a fixed baryon number during the different stages
of the evolution - the constraint given by the warm case with
trapped neutrinos for $g_4=0$ and for the cold beta equilibrium
case for $g_4\neq0$. In this way, the maximum mass and
radius of the star decrease with time. A separate analysis shows
that the mass and radius of the star increase when rotation is
included. A study of cooling effects in the first seconds of the
neutron star life together with rotation and
angular momentum conservation is currently under investigation.

\end{document}